\documentclass[12pt,a4paper]{article}

\usepackage{geometry}
\RequirePackage{amsmath}
\RequirePackage{mathrsfs}
\RequirePackage{amsfonts}
\RequirePackage{dsfont}
\RequirePackage{braket}
\RequirePackage{tabularx}
\usepackage{multirow}
\RequirePackage{nicefrac}
\RequirePackage{graphicx}
\RequirePackage{booktabs}
\RequirePackage{colortbl}
\RequirePackage{xcolor}
\RequirePackage[symbol]{footmisc}
\usepackage{mathtools}
\usepackage{comment}
\usepackage{blkarray}
\usepackage{stackengine}
\usepackage{float}
\usepackage{rotating}
\usepackage{hhline}
\usepackage{tikz}
\usepackage[section]{placeins}
\newcommand{\boxalign}[2][0.97\textwidth]{
  \par\noindent\tikzstyle{mybox} = [draw=black,inner sep=6pt]
  \begin{center}\begin{tikzpicture}
   \node [mybox] (box){%
    \begin{minipage}{#1}{\vspace{-5mm}#2}\end{minipage}
   };
  \end{tikzpicture}\end{center}
}
\def\AEF{A.E. Faraggi}

\def\IJMP#1#2#3{{\it Int.\ J.\ Mod.\ Phys.}\/ {\bf A#1} (#2) #3}

\def\EJP#1#2#3{{\it Eur.\ Phys.\ Jour.}\/ {\bf C#1} (#2) #3}

\def\JHEP#1#2#3{{\it JHEP}\/ {\bf #1} (#2) #3}
\def\NPB#1#2#3{{\it Nucl.\ Phys.}\/ {\bf B#1} (#2) #3}
\def\PLB#1#2#3{{\it Phys.\ Lett.}\/ {\bf B#1} (#2) #3}
\def\PRD#1#2#3{{\it Phys.\ Rev.}\/ {\bf D#1} (#2) #3}
\def\PRL#1#2#3{{\it Phys.\ Rev.\ Lett.}\/ {\bf #1} (#2) #3}
\def\PRT#1#2#3{{\it Phys.\ Rep.}\/ {\bf#1} (#2) #3}
\def\etal{{\it et al\/}}

\def\beq{\begin{equation}}
\def\eeq{\end{equation}}
\def\beqn{\begin{eqnarray}}
\def\eeqn{\end{eqnarray}}

\def\unahe{{${\overline{\rm NAHE}}$}}
\usepackage{empheq}
\usepackage[most]{tcolorbox}
\usepackage{blkarray}
\newtcbox{\mymath}[1][]{%
    nobeforeafter, math upper, tcbox raise base,
    enhanced, colframe=blue!30!black,
    colback=blue!30, boxrule=1pt,
    #1}

\newcommand{\CC}[2]{C{#1\atopwithdelims[]#2}}

\newcommand{\ba}{\begin{eqnarray}}
\newcommand{\ea}{\end{eqnarray}}
\DeclareRobustCommand{\sqbinom}{\genfrac[]{0pt}{}}

\numberwithin{equation}{section}

\begin{document}
\begin{titlepage}
\samepage{
\setcounter{page}{1}
\rightline{}
\rightline{October 2020}

\vfill
\begin{center}
  {\Large \bf{
  Type 0 $\mathbb{Z}_2\times \mathbb{Z}_2$ Heterotic String Orbifolds
        \\ \medskip
      and Misaligned Supersymmetry }}

\vspace{1cm}
\vfill

{\large Alon E. Faraggi\footnote{E-mail address: alon.faraggi@liv.ac.uk}, %\\
         Viktor G. Matyas\footnote{E-mail address: viktor.matyas@liv.ac.uk}
 and Benjamin Percival\footnote{E-mail address: benjamin.percival@liv.ac.uk} 
}
\\

\vspace{1cm}

{\it Dept.\ of Mathematical Sciences, University of Liverpool, Liverpool
L69 7ZL, UK\\}

\vspace{.025in}
\end{center}

\vfill
\begin{abstract}
\noindent

The $\mathbb{Z}_2\times \mathbb{Z}_2$ heterotic string orbifold yielded a large space of phenomenological
three generation models and serves as a testing ground to explore how the Standard Model
of particle physics may be incorporated in a theory of quantum gravity. In this paper we explore the existence of type 0 models in this class of string
compactifications. We demonstrate the existence of type 0
$\mathbb{Z}_2\times \mathbb{Z}_2$ heterotic string orbifolds, and show that there exist a large degree 
of redundancy in the space of GGSO projection coefficients when the type 0 restrictions
are implemented. We explore the existence of such configurations in several constructions. 
The first correspond to essentially a unique configuration 
out of a priori $2^{21}$ discrete GGSO choices. We demonstrate this uniqueness analytically, as well as by the corresponding analysis of the partition function.
A wider classification is performed in
$\tilde S$--models and $S$--models, where the first class correspond to compactifications
of a tachyonic ten dimensional heterotic string vacuum, whereas the second correspond to 
compactifications of the ten dimensional non--tachyonic $SO(16)\times SO(16)$. We show
that the type 0 models in both cases contain physical tachyons at the free fermionic point
in the moduli space. These vacua are therefore necessarily unstable, but may be instrumental in exploring the string dynamics in cosmological scenarios. 
we analyse the properties of the string one--loop amplitude.
Naturally, these are divergent due to the existence of tachyonic states.
We show that once the tachyonic states are removed by hand the amplitudes are
finite and exhibit a form of misaligned supersymmetry. 

\end{abstract}

\smallskip}

\end{titlepage}

\section{Introduction}\label{intro}

The holy grail of contemporary theoretical physics is
the synthesis of quantum mechanics and general relativity.
String theory provides an arena to explore this synthesis
within a perturbatively self--consistent framework. The
important advantage of string theory is that its internal
consistency conditions mandate the appearance of the gauge
and matter structures that exist in the observed world,
hence facilitating the development of a phenomenological
approach. It seems likely that there exist numerous
mathematical structures that can be used to
developed consistent theories of quantum gravity,
and that mathematical consistency alone will not
be sufficient for the development of a physically relevant
theory, {\it i.e.} one that is relevant for experimental
observations. It is therefore sensible to construct string models
that aim to reproduce the physical characteristics of the
observed physical world. Such toy models can in turn be
used to explore the mathematical structures underlying string
theory and their possible relation to observable phenomena.
Of particular interest is the relation of string vacua to their
effective field theory limits. In this respect it should be noted
that contemporary understanding of string theory is confined to
its static limits. Dynamical questions in nature, such as the
cosmological evolution in the early universe, are explored
at present primarily in the effective field theory limit.

Given the vast space of potential string vacua, it makes further
sense to study classes of models that exhibit certain common structures
and that produce examples of quasi--realistic phenomenological models.
While it is folly to expect at present any of the models to be fully realistic,
it is feasible that some characteristics of a given class of models may
be imprinted in the observable data. The task of string phenomenology is to
identify such imprints in different classes of string compactifications,
and how the experimental data may discern between them.
One class of string compactification that have been explored in
this spirit are the $\mathbb{Z}_2\times \mathbb{Z}_2$ heterotic string orbifolds.

The $\mathbb{Z}_2\times \mathbb{Z}_2$ heterotic string orbifolds have been studied in the
free fermionic formulation of the heterotic string \cite{fff}
since the late eighties. They produce a large space of
quasi--realistic three generation models with different
unbroken $SO(10)$ and $E_6$ subgroups
\cite{fsu5, fny, alr, slm, lrs, acfkr, su62, frs,
  slmclass, lrsclass, lrsfertile}. Quasi--realistic 
$\mathbb{Z}_2\times \mathbb{Z}_2$ orbifold models have also been constructed by using the
free bosonic formulation \cite{stefanetal}. The two formulations
provide complementary tools to explore the same physical spaces,
and detailed dictionaries have been developed to translate models
from one formulation to the other \cite{z2xz2}. The $\mathbb{Z}_2\times \mathbb{Z}_2$
orbifolds exhibit a rich symmetry structure, which has been the interest
of numerous mathematical studies \cite{mathematical}.
The rich symmetry structure of the $\mathbb{Z}_2$ orbifold may also be relevant
for the Standard Model flavour problem \cite{fh, nillesflavours}.
Indeed, it was demonstrated that quasi--realistic fermion mass and
mixing spectrum may be obtained from the fermionic $\mathbb{Z}_2\times \mathbb{Z}_2$
orbifolds \cite{fh}. The $\mathbb{Z}_2\times \mathbb{Z}_2$ orbifold compactifications
have also been studied in other string limits \cite{as}. 

While the majority of studies of phenomenological string models
have been of $N=1$ supersymmetric vacua, non--supersymmetric
compactifications have been of increased interest over the past
decade. The starting point for most of these non--supersymmetric
constructions is the $SO(16)\times SO(16)$ heterotic string in ten
dimensions \cite{dh, itoyama, nonsusy, aafs, ADM, FR}. However,
it is well known that string theory also admits solutions that are
tachyonic in ten dimensions \cite{dh, kltclas, gv}. It was recently
argued that these ten dimensional vacua may also serve as starting points for
the constructions of viable phenomenological models, provided that the
tachyonic modes are projected from the physical spectrum in the
phenomenological four dimensional models \cite{spwsp, stable, so10tclass}.
Indeed, a tachyon free three generation Standard--like Model
was constructed in this class \cite{stable},
whereas in ref. \cite{so10tclass} a broad classification of models
with unbroken $SO(10)$ gauge group was presented %ref. \cite{so10tclass},
including the analysis of their vacuum energy, and
numerous models that satisfy the condition $n_b=n_f$, {\it i.e.} models
with equal numbers of massless bosons and fermions.

In this paper, we extend the analysis of the aforementioned classification
of this class of vacua. Our interest here is in the existence of models
that do not contain any fermions at all. Such models are known in the
literature as type 0 models and have been of interest in other string theory
limits \cite{type0string}. Such models are of particular interest to explore
the boundaries of the space of $\mathbb{Z}_2\times \mathbb{Z}_2$ orbifold compactifications.
It is plausible that progress on some of the phenomenological
issues in string theory, in particular in relation to the
cosmological evolution and vacuum selection, will be obtained
by improved understanding of these vacua. Moreover, it is likely
that further insight
can be achieved by exploring some of the features of these vacua in
connection with the phenomenological string vacua.
In this paper we therefore pursue this line of investigation. We present
several type 0 models in this class. We further adapt the systematic
classification method that was developed using the free fermionic rules
\cite{gkr, fknr, fkr, acfkr, frs, slmclass, lrsclass, lrsfertile, so10tclass}
to this class of models. This requires careful analysis of tachyonic
states that proliferate
in these configurations. While we do not find any model which
is completely free of tachyonic states, we present a model with a minimal
set of tachyonic states. % that arise from Neveu--Schwarz sector. 
Similar to the
analysis in ref. \cite{FR}, we may explore the possibility that these
%untwisted 
tachyonic states become massive when the moduli is moved away from
the free fermionic point. %, while the twisted spectrum is unchanged.
Another issue that we analyse in detail is the calculation of the vacuum
energy and the finiteness properties of the string one--loop amplitude.
Naturally, these are divergent due to the existence of tachyonic states.
However, once the tachyonic states are removed by hand the amplitudes are
finite and exhibit a form of misaligned supersymmetry. 

Our paper is organised as follows: in Section \ref{model} we present an explicit example of a type 0 model that does not contain any massless fermionic states. 
    In Sections \ref{analyticConds}--\ref{equivalence} 
we show that the type 0 constraints are in fact very 
restrictive and showing that for the chosen set of basis vectors the type 0 model is 
in fact unique. Hence, we demonstrate the existence of a huge redundancy in the space of 
Generalised GSO phases that span the models. We show that the type 0 model in this
configuration does contain a tachyonic state that transform in the vector representation 
of the hidden sector gauge group. A question of interest that we undertake in the following 
sections is the existence of a tachyon free type 0 model. We investigate this question by adopting the free fermionic classification methodology to these models that enables the scan
of large space of models and the extraction of models with specific characteristics. We explore 
the existence of tachyon free type 0 models in both $\tilde S$--models as well as 
$S$--models that utilise the SUSY generating basis vector, Our scan yields a null result, 
which suggests that such a tachyon free model may only exist away from the free fermionic 
point. In Section \ref{MSSec} we demonstrate the existence of a misalligned supersymmetry
in the type 0 models that guarantee the finiteness of the one--loop amplitude, aside from 
the divergence due to the tachyonic states. Section \ref{conclude} concludes our paper.

\section{Type 0 $\mathbb{Z}_2\times \mathbb{Z}_2$ Heterotic String Orbifold}\label{model}

In the free fermionic contruction \cite{fff} models are specified
in terms of boundary condition basis vectors and one--loop Generalised
GSO phases. Details of the formalism are to be found in the literature
and are not repeated here. We adopt the conventional notation used in the
free fermionic constructions
\cite{fsu5, fny, alr, slm, lrs, gkr, acfkr, su62, frs, slmclass, lrsclass,
  lrsfertile}. The first model that we present uses the \unahe--set
that was introduced in \cite{spwsp, stable}. In this set the basis
vector $S$ that generates spacetime supersymmetry in NAHE--based models
\cite{nahe} is augmented with four periodic right--moving fermions,
which amounts to making the gravitinos massive. This introduces a general
$S\rightarrow {\tilde S}$ map in the space of models that was discussed in
detail in ref. \cite{so10tclass}. The set of basis vectors is given by
\begin{align}\label{basis}
\mathds{1}&=\{\psi^\mu,\
\chi^{1,\dots,6},y^{1,\dots,6}, w^{1,\dots,6}\ | \ \overline{y}^{1,\dots,6},\overline{w}^{1,\dots,6},
\overline{\psi}^{1,\dots,5},\overline{\eta}^{1,2,3},\overline{\phi}^{1,\dots,8}\},\nonumber\\
\tilde{S}&=\{{\psi^\mu},\chi^{1,\dots,6} \ | \ \overline{\phi}^{3,4,5,6}\},\nonumber\\
%{e_i}&=\{y^{i},w^{i}\; | \; \overline{y}^{i},\overline{w}^{i}\}, 
%\ \ \ i=1,...,6
%\nonumber\\
{b_1}&=\{\chi^{34},\chi^{56},y^{34},y^{56}\; | \; \overline{y}^{34},
\overline{y}^{56},\overline{\eta}^1,\overline{\psi}^{1,\dots,5}\},\\
{b_2}&=\{\chi^{12},\chi^{56},y^{12},w^{56}\; | \; \overline{y}^{12},
\overline{w}^{56},\overline{\eta}^2,\overline{\psi}^{1,\dots,5}\},\nonumber\\
{b_3}&=\{\chi^{12},\chi^{34},w^{12},w^{34}\; | \; \overline{w}^{12},
\overline{w}^{34},\overline{\eta}^3,\overline{\psi}^{1,\dots,5}\},\nonumber\\
z_1&=\{\overline{\phi}^{1,\dots,4}\},\nonumber\\
x&=\{\overline{\psi}^{1,\dots,5}, \overline{\eta}^{1,2,3}\}.\nonumber
\nonumber
\end{align}
A model may then be specified through the assignment of modular invariant GGSO phases $\CC{v_i}{v_j}$ between the basis vectors. A type 0 configuration arises for the assignment
{\begin{equation}
\small
\CC{v_i}{v_j}= 
\begin{blockarray}{ccccccccc}
&\mathbf{1}& \tilde{S} & b_1 & b_2&b_3&z_1 & x \\
\begin{block}{c(rrrrrrrr)}
\mathbf{1}& 1&  1& -1& -1& -1& -1& -1\ \\
\tilde{S} & 1& -1& -1& -1& -1&  1&  1\ \\
b_1       &-1&  1& -1& -1& -1&  1&  1\ \\
b_2       &-1&  1& -1& -1& -1&  1&  1\ \\
b_3       &-1&  1& -1& -1& -1&  1&  1\ \\
z_1       &-1& -1&  1&  1&  1& -1&  1\ \\ 
x         &-1&  1& -1& -1& -1&  1&  1\ \\ 
\end{block}
\end{blockarray}
\label{ggsophases}
\end{equation}}

All models in this basis will have gauge bosons arising from the Neveu--Schwarz (NS) sector that produce the
vector bosons of a $SO(10)\times U(1)^3\times SO(4)^3\times 
SU(2)^8$ gauge symmetry. Additional vector bosons may arise from the sectors
in the $z_1$,
$z_3={\bf 1}+{\tilde S}+b_1+b_2+b_3=\{{\bar\phi}^{1,2,7,8}\}$
and
$z_4={\bf 1}+{\tilde S}+b_1+b_2+b_3+z_1=\{{\bar\phi}^{3,4,7,8}\}$, which can affect the observable and the hidden gauge group factors or just the hidden, depending on the right-moving oscillator. A solely observable gauge enhancement may also arise from the sector $x$. In the above model, the hidden $SU(2)^8$ gauge symmetry
is enhanced to $SO(16)$ by vector bosons arising
in $z_1$, $z_3$ and $z_4$.
The four dimensional
gauge group is therefore
$$SO(10)\times U(1)^3\times SO(4)^3\times SO(16).$$
The NS--sector and the three enhancement sectors above produce in total
sixteen tachyonic states that transform in the 16 representation
of the hidden $SO(16)$ gauge symmetry. The vectorial fermionic sectors in the model are
\beqn
\label{VectFerms}
& &{\tilde S};\nonumber\\
& &{\tilde S} +z_1;\\
& &{\tilde S}+z_4\sim S+z_2,\nonumber 
\eeqn
while the massless fermionic spinorial sectors are
\beqn
\label{SpinFerms}
& &{\tilde S}+b_{1,2,3}+x;\nonumber\\
& &{\tilde S}+b_{1,2,3}+x+z_1;\\
& &{\tilde S}+b_{1,2,3}+x+z_4 ;\nonumber\\
& &{\tilde S}+z_3\sim S+z_1+z_2;\nonumber
\eeqn
where we defined $z_2=\{{\bar\phi}^{5, \cdots, 8}\}$
and $S=\{\psi^{1,2}, \chi^{1,\cdots, 6}\}$, neither of which are 
basis vector combinations in the additive group. 
We note that the $S$--vector coincides with the supersymmetry generator
in supersymmetric free fermionic models, as well as those that are
compactifications of the ten dimensional $SO(16)\times SO(16)$
heterotic string. 
These definitions comply with the terminology used in the classification
of the supersymmetric free fermionic heterotic string models.
We emphasise that the absence of the $S$--vector from the additive group is 
the crucial feature of the $\tilde S$--models. As discussed in refs. 
\cite{spwsp, stable, so10tclass} the absence of the $S$--vector is the characteristic 
property of vacua that descend from the tachyonic ten dimensional vacua. 
The massless states from all the fermionic sectors are projected out from
the physical spectrum by the choice of GGSO phases in eq. (\ref{ggsophases}).
In addition to the NS--sector, sectors giving rise to spacetime massless
bosonic states are
\beqn
& &x;\nonumber\\
& &z_{1,3,4};\nonumber\\
& &b_{1,2,3};\nonumber\\
& &b_{1,2,3}+x;\label{Bosons}\\
& &b_{1,2,3}+x+z_1;\nonumber\\
& &b_{1,2,3}+x+z_3;\nonumber\\
& &b_{1,2,3}+x+z_1+z_3.\nonumber
\eeqn
They give rise to scalar spacetime bosons
that transform in representations of the four dimensional gauge symmetry.
They are of no particular interest here and we do not list their detail explicitly. 

In Section \ref{StModels} and \ref{SModels} we perform a more general
search for similar type 0 heterotic string models using the
free fermionic classification methodology.
In particular, we search for type 0
models without tachyons. Our search is conducted
using the ${\tilde S}$ based models as well as models that use $S$. First, however, it is interesting to study a bit more closely the basis (\ref{basis}), and what additional constraints type 0 vacua may satisfy.

\subsection{Analytic conditions on type 0 vacua} \label{analyticConds}

Since we are interested in the construction and analysis of type 0 vacua, we focus on the massless
fermionic sectors and will seek to project them out, leaving only bosonic states at the massless level. The Hilbert space in the free fermionic construction is given by the collection of GGSO-projected states $\ket{S_\xi}$ 
\begin{equation}
    \mathcal{H}=\bigoplus_{\xi\in\Xi}\prod^{k}_{i=1}
\left\{ e^{i\pi v_i\cdot F_{\xi}}\ket{S_\xi}=\delta_{\xi}
\CC{\xi}{v_i}^*\ket{S_\xi}\right\}. 
\end{equation}
For the type 0 case, this will only have contributions from sectors $\xi$ in the additive space $\Xi$ with bosonic spin statistic index $\delta_{\xi}=+1$ at the massless level.
%, \textit{i.e.} when $\dfrac{\xi_L \cdot\xi_L}{8}+N_L=\dfrac{1}{2}$  and $\dfrac{\xi_R \cdot\xi_R}{8}+N_R=1$ with $N_L,N_R$ counting the left and right oscillator numbers, respectively. 

For the basis (\ref{basis}), These definitions comply with the terminology used in the classification
of the supersymmetric free fermionic heterotic string models. In particular, we note that the $S$--vector coincides with the supersymmetry generator
in supersymmetric free fermionic models, as well as those that are
compactifications of the ten dimensional $SO(16)\times SO(16)$
heterotic string which we analyse in Section \ref{SModels}. 

However, since we want to find choices of GGSO phases for which the fermionic sectors may be projected out to leave type 0 vacua, it is worth exploring explicitly the analytic conditions on their projection. Another important part of the analysis will be to consider the presence of tachyonic sectors in our models which we turn to in Section \ref{tachAnalysis}. Due to our basis being relatively simple, writing down analytic conditions for generating type 0 models seems tractable \textit{a priori} and we will see that, indeed, it is entirely solvable.

In order to project the fermionic massless sectors given in equations (\ref{VectFerms}) and (\ref{SpinFerms}) we can write down analytic conditions from the GGSO projection equation for the existence of type 0 models. For the fermionic vectorial sectors (\ref{VectFerms})
\begin{align}
\{\bar{y},\bar{w}\}\ket{\tilde{S}} \text{ projected } 
\iff & \left( 1-\CC{\tilde{S}}{x}\right)\left( 1-\CC{\tilde{S}}{z_3}\right)=0 \label{VF1} \\
\{\bar{\psi}^{1,...,5},\bar{\eta}^{1,2,3}\}\ket{\tilde{S}} \text{ projected }  
\iff & \left( 1+\CC{\tilde{S}}{x}\right)\left( 1-\CC{\tilde{S}}{z_3}\right)=0 \label{VF2} \\
\{\bar{\phi}^{1,2,7,8}\}\ket{\tilde{S}} \text{ projected }
\iff & \left( 1-\CC{\tilde{S}}{x}\right)\left( 1+\CC{\tilde{S}}{z_3}\right)=0 \label{VF3} \\
\{\bar{y},\bar{w}\}\ket{\tilde{S}+z_1} \text{ projected } \iff & \left( 1-\CC{\tilde{S}+z_1}{x}\right)\left( 1-\CC{\tilde{S}+z_1}{z_4}\right)=0 \label{VF4} \\
\{\bar{\psi}^{1,...,5},\bar{\eta}^{1,2,3}\}\ket{\tilde{S}+z_1} \text{ projected } 
\iff & \left( 1+\CC{\tilde{S}+z_1}{x}\right)\left( 1-\CC{\tilde{S}+z_1}{z_4}\right)=0 \label{VF5} \\
\{\bar{\phi}^{3,4,7,8}\}\ket{\tilde{S}+z_1} \text{ projected } 
\iff & \left( 1-\CC{\tilde{S}+z_1}{x}\right)\left( 1+\CC{\tilde{S}+z_1}{z_4}\right)=0 \label{VF6} \\
\{\bar{y},\bar{w}\}\ket{\tilde{S}+z_4} \text{ projected } 
\iff & \left( 1-\CC{\tilde{S}+z_4}{x}\right)\left( 1-\CC{\tilde{S}+z_4}{z_1}\right)=0 \label{VF7} 
\end{align}
\begin{align}
\{\bar{\psi}^{1,...,5},\bar{\eta}^{1,2,3}\}\ket{\tilde{S}+z_4} \text{ projected } \iff & \left( 1+\CC{\tilde{S}+z_4}{x}\right)\left( 1-\CC{\tilde{S}+z_4}{z_1}\right)=0 \label{VF8} \\
\{\bar{\phi}^{1,2,3,4}\}\ket{\tilde{S}+z_4} \text{ projected }
\iff & \left( 1-\CC{\tilde{S}+z_4}{x}\right)\left( 1+\CC{\tilde{S}+z_4}{z_1}\right)=0 \label{VF9} 
\end{align}
Using the ABK rules on equations (\ref{VF1}), (\ref{VF2}) and (\ref{VF3}) we deduce that
 \begin{equation}\label{StConds}
        \CC{\tilde{S}}{x}=1 \ \text{ and }  \ \CC{\tilde{S}}{b_1}\CC{\tilde{S}}{b_2}\CC{\tilde{S}}{b_3}=-1
\end{equation}
using these results and the ABK rules on equations (\ref{VF4}), (\ref{VF5}) and (\ref{VF6}) gives the further results 
\begin{equation}\label{z1Conds}
        \CC{z_1}{x}=1 \ \text{ and }  \ \CC{z_1}{b_1}\CC{z_1}{b_2}\CC{z_1}{b_3}=1.
\end{equation}
Finally, the first bracket of equations (\ref{VF7}), (\ref{VF8}) and (\ref{VF9}) implies the result
\begin{equation}\label{xConds}
         \CC{x}{1}\CC{x}{b_1}\CC{x}{b_2}\CC{x}{b_3}=1.
\end{equation}
Meanwhile, for the fermionic spinorial sectors (\ref{SpinFerms}) we have
\begin{align}
&\tilde{S}+b_i+x \text{ projected }\nonumber \\ 
\iff & \left( 1-\CC{\tilde{S}+b_i+x}{b_j+b_k+x}\right)\left( 1-\CC{\tilde{S}+b_i+x}{z_3}\right)=0 \label{SF1} \\
&\tilde{S}+b_i+z_1+x \text{ projected }\nonumber \\ 
\iff & \left( 1-\CC{\tilde{S}+b_i+z_1+x}{b_j+b_k+x}\right)\left( 1-\CC{\tilde{S}+b_i+z_1+x}{z_4}\right)=0 \label{SF2}  \\
&\tilde{S}+b_i+z_4+x \text{ projected } \nonumber\\ 
\iff & \left( 1-\CC{\tilde{S}+b_i+z_4+x}{b_j+b_k+x}\right)\left( 1-\CC{\tilde{S}+b_i+z_4+x}{z_1}\right)=0  \label{SF3} \\
&\tilde{S}+z_3 \text{ projected } \nonumber\\ 
\iff & \left( 1-\CC{\tilde{S}+z_3}{b_1+b_2+b_3+x}\right)\left( 1-\CC{\tilde{S}+z_3}{x}\right)=0  \label{SF4}
\end{align}
where $i\neq j \neq k=1,2,3$. Using results (\ref{StConds}),(\ref{z1Conds}) and (\ref{xConds}) in equation (\ref{SF3}) implies that  
\begin{equation}\label{z1Fixed}
\CC{z_1}{b_1}=\CC{z_1}{b_2}=\CC{z_1}{b_3}=1
\end{equation}
and using results (\ref{StConds}) and (\ref{xConds}) in equation (\ref{SF1}) allows us to deduce the results
\begin{equation}\label{bibjConds}
 \CC{\tilde{S}}{b_1}\CC{b_1}{b_2}\CC{b_1}{b_3}=-1, \ \ \ \CC{\tilde{S}}{b_2}\CC{b_2}{b_1}\CC{b_2}{b_3}=-1 \ \ \ \text{and} \ \ \ \CC{\tilde{S}}{b_3}\CC{b_3}{b_1}\CC{b_3}{b_2}=-1
\end{equation}
since (\ref{StConds}) means there are only two independent equations here, we can use this result to fix  two of the phases: $\CC{b_1}{b_2},\CC{b_1}{b_3}$ and $\CC{b_2}{b_3}$. 

Gathering together the results (\ref{StConds}), (\ref{z1Conds}), (\ref{xConds}), (\ref{z1Fixed}) and (\ref{bibjConds}) we find the following necessary and sufficient conditions on the projection of massless fermions within models derived from the basis (\ref{basis})
%\begin{equation}
\boxalign{\begin{align}
%\begin{array}{rcl}
\label{Type0Conds}
 \CC{\tilde{S}}{x}&=1, \ \ \ \CC{z_1}{x}=1,  \ \ \  \CC{z_1}{b_1}=\CC{z_1}{b_2}=\CC{z_1}{b_3}=1 \nonumber\\ 
 \CC{\tilde{S}}{b_1}&=-\CC{\tilde{S}}{b_2}\CC{\tilde{S}}{b_3} \nonumber\\ 
  \CC{x}{1}&=~~\CC{x}{b_1}\CC{x}{b_2}\CC{x}{b_3} \\ 
  \CC{b_2}{b_3}&=-\CC{\tilde{S}}{b_2}\CC{b_1}{b_2}\nonumber\\
  \CC{b_3}{b_1}&=-\CC{\tilde{S}}{b_2}\CC{\tilde{S}}{b_3}\CC{b_1}{b_2}\nonumber.
%\end{array}
\end{align}}
%\end{equation}
These conditions mean that 9 of the 21 GGSO phases are fixed in order to obtain type 0 vacua. Hence, the number of possible type 0 models is reduced to $2^{12}=4096$. 

Armed with the conditions (\ref{Type0Conds}) we can look now at the bosonic sectors (\ref{Bosons}) and in fact prove that all these sectors must appear in all 4096 possible type 0 models. To prove this we can go through the projection conditions as we did above for the fermionic sectors. In particular, taking the sectors $b_i$, for $i=1,2,3$ we get 
\begin{equation}
b_i \text{ survives }
\iff \CC{b_i}{\tilde{S}+b_j+b_k}=1 \ \ \text{ and } \ \ \CC{b_i}{z_1} =1 \label{B1} 
\end{equation}
these conditions coincide exactly with conditions (\ref{bibjConds}) and (\ref{z1Fixed}), respectively, from the projection of fermions analysis above. A similar result can easily be found for the other spinorial bosonic sectors: $b_{1,2,3}+x+z_1,b_{1,2,3}+x+z_3$ and $b_{1,2,3}+x+z_4$. 

For the bosonic vectorial sectors: $b_i+x$, $i=1,2,3$ we have the conditions
\begin{align}
\{\bar{y},\bar{w},\bar{\psi}^{1,...,5},\bar{\eta}^{i},\bar{\phi}^{5,6}\}\ket{b_i+x} \text{ survives } 
\implies & \frac{1}{4}\left( 1+\CC{b_i+x}{z_3}\right)\left( 1+\CC{b_i+x}{z_1}\right)=1 \label{B2} \\
\{\bar{\phi}^{1,2}\}\ket{b_i+x} \text{ survives } 
\implies & \frac{1}{4}\left( 1-\CC{b_i+x}{z_3}\right)\left( 1-\CC{b_i+x}{z_1}\right)=1 \label{B3} \\
\{\bar{\phi}^{3,4}\}\ket{b_i+x} \text{ survives } 
\implies & \frac{1}{4}\left( 1+\CC{b_i+x}{z_3}\right)\left( 1-\CC{b_i+x}{z_1}\right)=1 \label{B4} \\
\{\bar{\phi}^{7,8}\}\ket{b_i+x} \text{ survives } 
\implies & \frac{1}{4}\left( 1-\CC{b_i+x}{z_3}\right)\left( 1+\CC{b_i+x}{z_1}\right)=1 \label{B5} 
\end{align}
the type 0 conditions (\ref{Type0Conds}) guarantee that $\CC{b_i+x}{z_3}=1$ and $\CC{b_i+x}{z_1}=1$ and thus the first case survives. 

Finally let us show that the bosonic sector $x$ survives. In order to make this sector massless there must be a left moving oscillator, which could make it a gauge boson if this oscillator is $\psi^\mu$. However, since for type 0 models $\CC{x}{\tilde{S}}=1$ only the states of the type $\{y^i/w^i\}_{1/2}\ket{x}$ can survive, which they must do due to $\CC{x}{z_1}=\CC{x}{z_3}=1$. 

For the $z_1$ massless sector the conditions for type 0 models necessitate that states of the type $\{y/w\}\{\bar{y}/\bar{w}\}\ket{z_1}$ and extra gauge bosons of the type $\{ \psi^\mu\}\{\bar{\phi}^{5,6,7,8}\}\ket{z_1}$ survive. Similarly for the $z_3$ massless sector type 0 models must have states of the type $\{y/w\}\{\bar{y}/\bar{w}\}\ket{z_3}$ and extra gauge bosons of the type $\{ \psi^\mu\}\{\bar{\phi}^{3,4,5,6}\}\ket{z_3}$. Finally, for the $z_4$ massless sector type 0 models must have states of the type $\{y/w\}\{\bar{y}/\bar{w}\}\ket{z_4}$ and extra gauge bosons of the type $\{ \psi^\mu\}\{\bar{\phi}^{1,2,5,6}\}\ket{z_4}$. Therefore, all type 0 models derived from this basis (\ref{basis}) have a hidden sector enhancement of $SU(2)^4\rightarrow SO(16)$.

This analysis tells us that all 4096 possible type 0 models contain all bosonic sectors \ref{Bosons} with the specific set of oscillators given above. In other words, their massless spectra are identical. Doing the counting of all the bosonic states can be shown to give 4264, which is thus the constant term in the $q$-expansion of the partition function in all 4096 cases. Having seen how restrictive the type 0 conditions $\ref{Type0Conds}$ are at the massless level it makes sense to analyse what happens with the tachyonic sectors for our type 0 models. 

\subsection{Tachyon analysis for type 0 vacua}\label{tachAnalysis}
The tachyonic sectors for models derived from the basis (\ref{basis}) come from the sectors $\ket{z_1}, \ \ket{z_3}$ and $\ket{z_4}$ as well as the untwisted tachyon of the NS sector. We can immediately see that all vacua in this basis will contain an untwisted tachyon $\{\bar{\phi}^{5,6}\}\ket{NS}$. This can be seen as being related to the absence of $z_2=\{\bar{\phi}^{5,6,7,8}\}$ in the basis which would allow for the projection of this tachyon since we would be equipped with the GGSO projection with the phase $\CC{NS}{z_2}=1$. 

In regard to the tachyons from the sectors $\ket{z_1}, \ \ket{z_3}$ and $\ket{z_4}$, we see that the type 0 conditions (\ref{Type0Conds}) necessitate their presence. For example, all phases that could project the $z_1$ tachyon: $\CC{z_1}{b_1},\CC{z_1}{b_2}, \CC{z_1}{b_3}, \CC{z_1}{x}$ are all equal to +1 and thus leave it in the Hilbert space. Therefore, we conclude that all type 0 in this construction contain the tachyons from the sectors $\ket{z_1}, \ \ket{z_3}$ and $\ket{z_4}$, along with the model-independent untwisted tachyon  $\bar{\phi}^{5,6}\ket{NS}$. 

\subsection{Equivalence of type 0 models}\label{equivalence}
Having shown that the massless spectrum and tachyonic sectors are identical for all the 4096 choices of GGSO phases consistent with the type 0 conditions (\ref{Type0Conds}), we might wonder whether these models are in fact identical at all massive levels. Calculating the partition function for all 4096 type 0 models proved that they indeed all have the same partition function 
\begin{equation}\label{PF}
     Z = 2\bar{q}^{-1} + 16q^{-1/2}\bar{q}^{-1/2} + 4264 + 45056 q^{1/4}\bar{q}^{1/4} + \cdots
\end{equation}
and thus there is only one type 0 model in our construction with degeneracy 4096.
This is a good example of the non-uniqueness
of the free fermionic construction definition of a model since the partition function (\ref{PF}) is invariant under the 12 phases: $\CC{1}{\tilde{S}}$, $\CC{1}{b_1}$, $\CC{1}{b_2}$, $\CC{1}{b_3}$, $\CC{1}{z_1}$, $\CC{\tilde{S}}{b_2}$, $\CC{\tilde{S}}{b_3}$, $\CC{\tilde{S}}{z_1}$, $\CC{b_1}{b_2}$, $\CC{x}{b_1}$, $\CC{x}{b_2}$, $\CC{x}{b_3}$. This result will ultimately be related to the many symmetries underlying models defined by the basis (\ref{basis}).

\section{Classification of Type 0 $\tilde{S}$-Models}\label{StModels}
Having found that type 0, tachyonic models exist for the simple basis (\ref{basis}), we can consider a more general basis
\begin{align}\label{basisStTi}
\mathds{1}&=\{\psi^\mu,\
\chi^{1,\dots,6},y^{1,\dots,6}, w^{1,\dots,6}\ | \ \overline{y}^{1,\dots,6},\overline{w}^{1,\dots,6},
\overline{\psi}^{1,\dots,5},\overline{\eta}^{1,2,3},\overline{\phi}^{1,\dots,8}\},\nonumber\\
\tilde{S}&=\{{\psi^\mu},\chi^{1,\dots,6} \ | \ \overline{\phi}^{3,4,5,6}\},\nonumber\\
{T_1}&=\{y^{1,2},w^{1,2}\; | \; \overline{y}^{1,2},\overline{w}^{1,2}\},\nonumber\\ 
{T_2}&=\{y^{3,4},w^{3,4}\; | \; \overline{y}^{3,4},\overline{w}^{3,4}\},\nonumber\\ 
{T_3}&=\{y^{5,6},w^{5,6}\; | \; \overline{y}^{5,6},\overline{w}^{5,6}\},\nonumber\\ 
%\nonumber\\
{b_1}&=\{\chi^{34},\chi^{56},y^{34},y^{56}\; | \; \overline{y}^{34},
\overline{y}^{56},\overline{\eta}^1,\overline{\psi}^{1,\dots,5}\},\\
{b_2}&=\{\chi^{12},\chi^{56},y^{12},w^{56}\; | \; \overline{y}^{12},
\overline{w}^{56},\overline{\eta}^2,\overline{\psi}^{1,\dots,5}\},\nonumber\\
{b_3}&=\{\chi^{12},\chi^{34},w^{12},w^{34}\; | \; \overline{w}^{12},
\overline{w}^{34},\overline{\eta}^3,\overline{\psi}^{1,\dots,5}\},\nonumber\\
z_1&=\{\overline{\phi}^{1,\dots,4}\},\nonumber
\nonumber
\end{align}
where the $T_i$, $i=1,2,3$ allow for internal symmetric shift in the compactified coordinates around the 3 tori. The only other difference to the basis (\ref{basis}) is that $x$ is now a linear combination: 
\begin{equation}
    x=b_1+b_2+b_3+T_1+T_2+T_3
\end{equation}
and we have the same combinations $z_3=1+\tilde{S}+b_1+b_2+b_3$ and $z_4=z_3+z_1$. We can further note that the space of independent GGSO phase configuration is now $2^{36}\sim 6 \times 10^{10}$ for this basis.

The addition of the $T_i$'s has some key consequences in relation to finding tachyon-free type 0 vacua. It multiplies the number of massless fermionic sectors and also increases the number of ways to project the (fermionic) sectors. Furthermore, we now have 15 tachyonic sectors: $z_1,z_3,z_4,T_i,z_1+T_i, z_3+T_i$ and $z_4+T_i$, $i=1,2,3$ rather than just the 3 for basis (\ref{basis}). We can notice that the model--independent NS tachyon $\{\bar{\phi}^{5,6}\}\ket{NS}$ remains in this construction so the minimal number of tachyons is to only have the NS tachyon.
\subsection{Fermionic sector analysis}
Using a similar methodology to Section \ref{analyticConds}, we wish to analyse the conditions on the projection of all fermionic sectors from these models. Due to increased size of the space of models from the added complexity of having of $T_{i=1,2,3}$ in the basis, we developed a computer algorithm to scan efficiently over the space of vacua and check for the absence of fermionic massless states.

We note that massless fermionic vectorials in these models arise from the sectors
\beqn
\label{VectFermsTi}
& &{\tilde S};\nonumber\\
& &{\tilde S} +z_1;\\
& &{\tilde S}+z_4\nonumber 
\eeqn
and the massless fermionic spinorial sectors from
\beqn
\label{SpinFermsTi}
& &{\tilde S}+b_i+b_j+T_k+pT_i+qT_j;\nonumber\\
& &{\tilde S}+b_i+b_j+z_1+T_k+pT_i+qT_j;\\
& &{\tilde S}+b_{1,2,3}+x+z_4 ;\nonumber\\
& &{\tilde S}+z_3.\nonumber
\eeqn
Our computer algorithm can then be further applied to analyse the tachyonic sectors arising in type 0 models. The results of this computerised scan are presented in the following section.

\subsection{Results of classification}
\begin{figure}[!htb]
\centering
\includegraphics[width=0.8\linewidth]{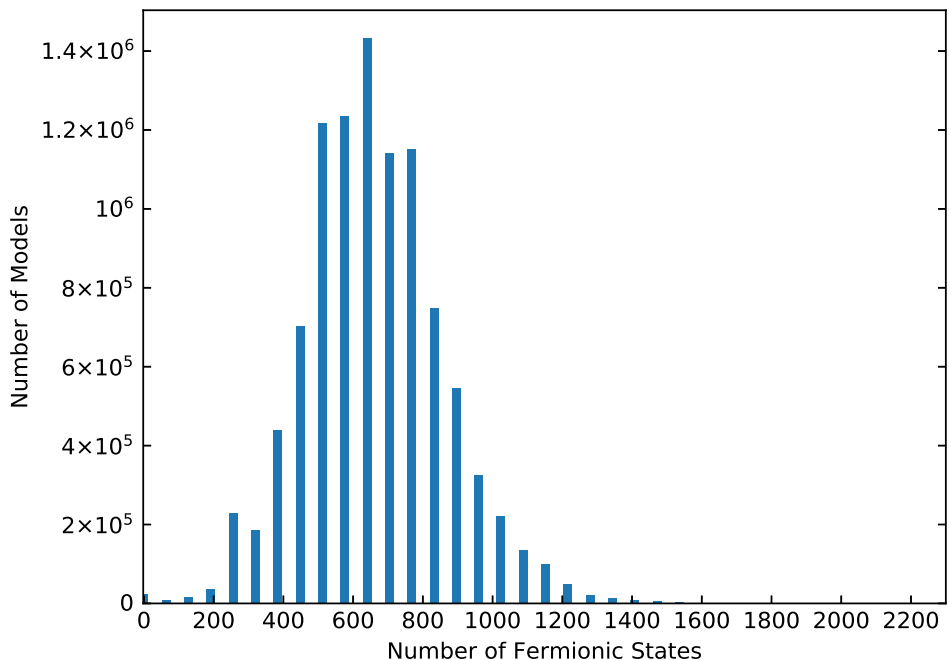}
\caption{\label{StNumFerms}\emph{Frequency plot for the number of fermionic states in a model from a sample of $10^7$ randomly generated GGSO configurations.}}
\end{figure}

By implementing the projection conditions on the massless fermionic sectors (\ref{VectFermsTi}) and (\ref{SpinFermsTi}) in a computer scan we can collect data for the number of fermionic states remaining in the Hilbert Space of a model and see how many are fermion-free and thus type 0. The distribution of the number of fermionic states for a scan of $10^7$ is displayed in Figure \ref{StNumFerms}. In this sample we find a total of 24508 which are free of fermionic states.

In order to gather a slightly larger sample of type 0 models in this basis we take a larger sample of $10^8$ models which still does not take much computing time. From this sample, we find 245685 type 0 models which gives a probability $\sim 2.46\times 10^{-3}$ for type 0 vacua in the total space. We now wish to classify these type 0 configurations according to which tachyonic sectors remain in their spectra (along with the model-independent untwisted tachyon), as shown in Table \ref{StTachs}.
\begin{table}[!htb]
\centering
\begin{tabular}{|c|c|c|c|}
\hline
&&&\\[-1.6ex]
$z_k$ Tachyon & $z_k+T_i$ Tachyon &$\{\bar{\lambda}^a\}\ket{T_i}$ Tachyon &Frequency\\[0.8ex]
\hline
0& 2& 2&    42773\\ \hline
1& 2& 1&   33513\\ \hline
1& 2& 2&    19402\\ \hline
1& 0& 2&    17405\\ \hline
1& 0& 1&   17140\\ \hline
1& 1& 2&  12056\\ \hline
0& 3& 1&    11996\\ \hline
3& 0& 1&     7141\\ \hline
0& 1& 3&     6044\\ \hline
3& 0& 2&     5708\\ \hline
1& 2& 3&     5575\\ \hline
1& 2& 0&     5175\\ \hline
1& 1& 1&     5170\\ \hline
1& 4& 2&     5071\\ \hline
0& 4& 2&     5017\\ \hline
0& 0& 2&     4262\\ \hline
0& 2& 3&     4253\\ \hline
1& 4& 1&     4226\\ \hline
3& 0& 0&     3827\\ \hline
0& 3& 2&     3405\\ \hline
0& 1& 1&     3389\\ \hline
1& 4& 0&    3322\\ \hline
1& 1& 3&     2625\\ \hline
1& 3& 3&     2179\\ \hline
0& 3& 3&     1774\\ \hline
0& 4& 3&     1724\\ \hline
3& 3& 2&     1713\\ \hline
1& 3& 2&     1631\\ \hline
3& 0& 3&     1529\\ \hline
3& 3& 3&     1168\\ \hline
1& 5& 2&      913\\ \hline
1& 5& 3&      888\\ \hline
0& 4& 1&      854\\ \hline
1& 0& 3&      840\\ \hline
1& 4& 3&      795\\ \hline
1& 6& 3&      583\\ \hline
0& 0& 3&      308\\ \hline
3& 6& 3&      291\\ \hline
\end{tabular}
\caption{\label{StTachs} \emph{Number of tachyonic sectors for 245685 type 0 $\tilde{S}$-models, where $k=1,3,4$, $i=1,2,3$ and $\bar{\lambda}^a$ is any right-moving oscillator with NS boundary condition.}}
\end{table}

These results clearly show that all type 0 models have both the model-independent untwisted tachyon and some combination of at least 2 twisted tachyonic sectors. One might wonder how general this result is since our sample size of $10^8$ only covers about $1:687$ models in the total space of GGSO phase configurations. Recalling the 4096 degeneracy factor from the analysis of models in the basis (\ref{basis}), we can reasonably suppose that type 0 models are highly constrained and degenerate also in the current construction where the $T_{i=1,2,3}$ are incorporated in the basis. 

\begin{figure}[t]
\centering
\hspace{1cm}
\includegraphics[width=0.85\linewidth]{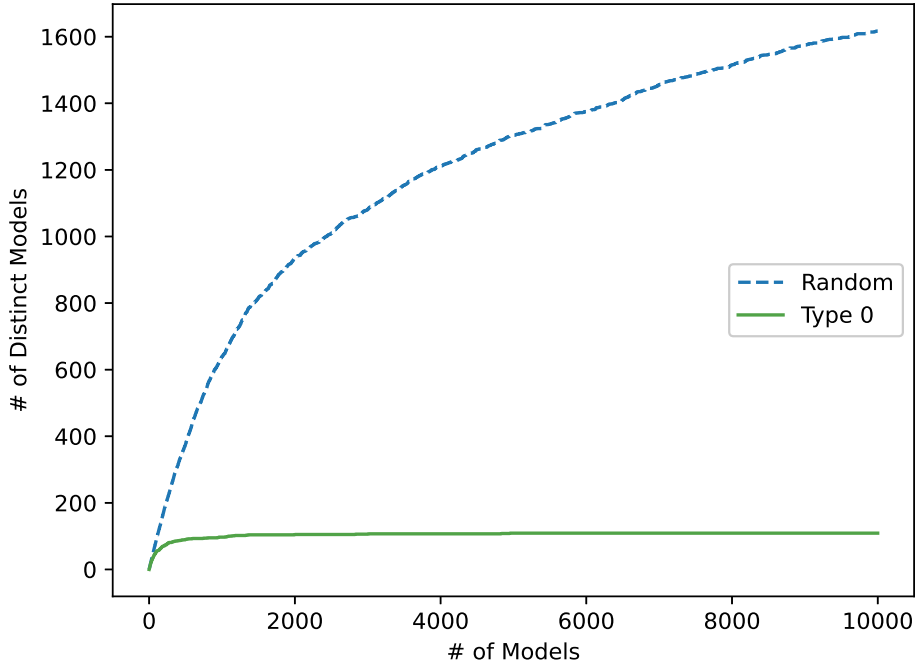}
\caption{\emph{The degeneracy of models for a random sample of models versus type 0 models for a sample of $10^4$ models each. We see that the space of type 0 models is indeed highly degenerate. }}
\label{Deg}
\end{figure}

To see this we took $10^4$ type 0 models out from the 245685 total sample and calculated their partition functions and found a total of 109 distinct ones. In Figure \ref{Deg} a comparison between the degeneracy of these $10^4$ type 0 models and those of a random sample of $10^4$ models is shown and the number of different type 0 models are seen to converge fast to just over 100, This shows, just as in the earlier case, that the subspace of type 0 vacua is highly symmetric. This result strongly justifies the generality of our results from the $10^8$ sample for the tachyonic analysis and makes it highly likely that our 245685 type 0 models from the $10^8$ sample captures all such unique models. In Section \ref{MSSec}, we will further discuss the structure of these type 0 models from the point of view of the partition function and one-loop vacuum energy.

\section{Classification of Type 0 $S$--Models}\label{SModels}
Having explored a space of $\tilde{S}$-models in the previous section we now wish to do the same analysis for models deriving from the ten-dimensional $SO(16)\times SO(16)$ non-supersymmetric heterotic string, which we will refer to as S-models since their basis contains the SUSY--generating vector $S$. The precise basis we use is: 
\begin{align}\label{basisS}
\mathds{1}&=\{\psi^\mu,\
\chi^{1,\dots,6},y^{1,\dots,6}, w^{1,\dots,6}\ | \ \overline{y}^{1,\dots,6},\overline{w}^{1,\dots,6},
\overline{\psi}^{1,\dots,5},\overline{\eta}^{1,2,3},\overline{\phi}^{1,\dots,8}\},\nonumber\\
S&=\{{\psi^\mu},\chi^{1,\dots,6} \},\nonumber\\
{T_1}&=\{y^{1,2},w^{1,2}\; | \; \overline{y}^{1,2},\overline{w}^{1,2}\},\nonumber\\ 
{T_2}&=\{y^{3,4},w^{3,4}\; | \; \overline{y}^{3,4},\overline{w}^{3,4}\},\nonumber\\ 
{T_3}&=\{y^{5,6},w^{5,6}\; | \; \overline{y}^{5,6},\overline{w}^{5,6}\},\nonumber\\ 
%\nonumber\\
{b_1}&=\{\chi^{34},\chi^{56},y^{34},y^{56}\; | \; \overline{y}^{34},
\overline{y}^{56},\overline{\eta}^1,\overline{\psi}^{1,\dots,5}\},\\
{b_2}&=\{\chi^{12},\chi^{56},y^{12},y^{56}\; | \; \overline{y}^{12},
\overline{y}^{56},\overline{\eta}^2,\overline{\psi}^{1,\dots,5}\},\nonumber\\
z_1&=\{\overline{\phi}^{1,\dots,4}\},\nonumber\\
z_2&=\{\overline{\phi}^{5,\dots,8}\},\nonumber
\nonumber
\end{align}
which is in fact identical to that used in ref. \cite{FR} and the same as that used in the supersymmetric classifications of \cite{gkr, fknr, fkr, acfkr, frs, slmclass, lrsclass, lrsfertile, so10tclass} up to the swap $T_{1,2,3}\rightarrow e_{1,...,6}$. 
As noted in these works, we will make regular use of the important linear 
combination $x$ \cite{xmap}, which appears as the combination
\begin{equation}
    x=1+S+\sum_{i=1}^3T_i+z_1+z_2
\end{equation}
in this basis and we further have the combination $b_3=b_1+b_2+x$ to give the generator of the third orbifold plane.

The untwisted gauge bosons in this construction generate a gauge group of $U(1)^6\times SO(10)\times U(1)^3 \times SO(8)^2$ and the full space of models is again given by the combinations of modular invariant GGSO phase configurations $2^{36}\sim 6\times 10^{10}$. % +1 in F&R is from [1][1] phase I guess... I always set this to +1... I believed this was a convention thing but maybe just for susy case... does this matter?

An important difference from the $\tilde{S}$-construction is that we do not have a model-independent tachyon as the NS tachyon is automatically projected. This leaves the door open for possible tachyon-free type 0 models.

Now we turn to the massless fermionic vectorial sectors which for our $S$-models arise from
\beqn
\label{SVectFerms}
& &S\nonumber\\
& &S+z_1 ;\\
& &S+z_2\nonumber\\
& & S+b_i+x+pT_j+qT_k\nonumber
\eeqn
and the massless fermionic spinorial sectors from
\beqn
\label{SSpinFerms}
& &S+x\nonumber\\
& &S+z_1+z_2\nonumber\\
& &S+b_i+pT_j+qT_k;\\
& &S+b_i+z_1+x+pT_j+qT_k ;\nonumber\\
& &S+b_i+z_2+x+pT_j+qT_k;\nonumber
\eeqn
where $i\neq j \neq k\in \{1,2,3\}$ and $p,q\in \{0,1\}$. We note that there are more fermionic sectors in this construction than in the $\tilde{S}$ case. In particular, the familiar $\mathbf{16}/\overline{\mathbf{16}}$ of $SO(10)$ from the sectors $S+b_i+pT_j+qT_k$ and vectorial $\mathbf{10}$ of $SO(10)$ from the sectors $(S+)b_i+x+pT_j+qT_k$ arise in this construction. However, for the $\tilde{S}$-models they were deliberately chosen to be absent and there are in fact no spinorial fermionic sectors with non-trivial representation under the $SO(10)$ observable gauge group factor.

As in the case of the $\tilde{S}$-models we use a computer algorithm to scan for type 0 configurations where these fermionic sectors are projected out. The results are presented in the following section.

\subsection{Results of classification}
\begin{center}
\begin{figure}[!htb]
\centering
\includegraphics[width=0.8\linewidth]{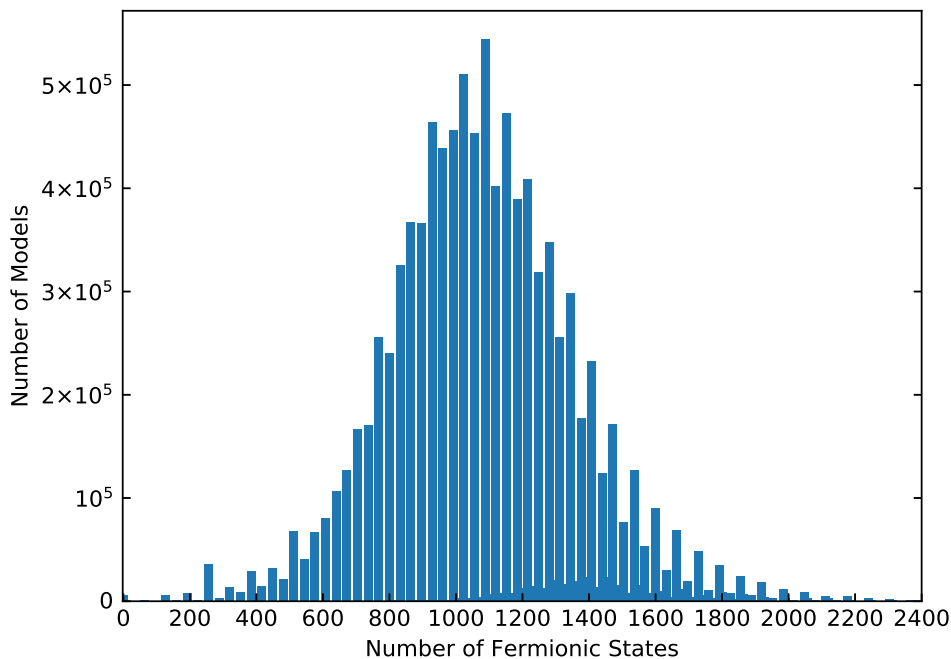}\caption{\label{NumFermsS}\emph{Frequency plot for the number of fermionic states for $S$-models from a random sample of $10^7$ GGSO configurations.}}
\end{figure}
\end{center}

As in the case of the $\tilde{S}$-models we generate a distribution for the number of fermionic states across a sample of $10^7$ randomly generated GGSO phase configurations. This is shown in Figure \ref{NumFermsS}. Comparing this to Figure \ref{StNumFerms} for the $\tilde{S}$ case we see a more dense distribution with more possible values for the fermionic states. 

As in the $\tilde{S}$ case we generate a larger sample of type 0 models by taking a scan of $10^8$ GGSO configurations. From this scan we find 54590 type 0 models which is probability $\sim 5.46\times 10^{-4}$ which makes them approximately 5 times rarer than in the $\tilde{S}$ case. The likely reason for this and main difference in general between the $S$ case and $\tilde{S}$ case is the already mentioned fact that in the $S$-models we have more fermionic massless sectors to project.  

Despite being rarer we already noted that we do not have any model-independent tachyons for these $S$-models which leaves the possibility of tachyon-free type 0 vacua open. The data for the numbers of tachyons is shown in Table \ref{STachs} and we see again that no tachyon-free models exist and that the minimal number of tachyonic sectors is 2, which always includes at least 1 vectorial tachyon of the type $\{\bar{\lambda}^a\}\ket{T_i}$.

\begin{table}[!htb]
\centering
\begin{tabular}{|c|c|c|c|}
\hline
&&&\\[-1.6ex]
$z_k$ Tachyon & $z_k+T_i$ Tachyon &$\{\bar{\lambda}^a\}\ket{T_i}$ Tachyon &Frequency\\[0.8ex]
\hline
1& 1& 2&    11605 \\ \hline
1& 0& 2 &   10471\\ \hline
1& 1& 1 &   4431\\ \hline
1& 0& 3 &  4388\\ \hline
2& 0& 2 &  4066\\ \hline
2& 0& 1 &   3749\\ \hline
1& 0& 1 &    3363\\ \hline
1& 1& 3 &    2384\\ \hline
1& 2& 2 &   1870\\ \hline
2& 0& 3 &    1509\\ \hline
1& 2& 3 &    1318\\ \hline
1& 2& 1 &    1080\\ \hline
2& 2& 1 &     871\\ \hline
2& 2& 2 &     538\\ \hline
0& 1& 3 &     488\\ \hline
0& 1& 2 &     454\\ \hline
0& 2& 1 &     299\\ \hline
1& 3& 1 &     291\\ \hline
1& 3& 2 &     290\\ \hline
0& 0& 2 &    236\\ \hline
1& 3& 3 &     189\\ \hline
0& 4& 3 &     151\\ \hline
0& 2& 3 &     151\\ \hline
0& 2& 2 &     135\\ \hline
0& 0& 3 &     135\\ \hline
2& 4& 1 &    128\\ \hline
\end{tabular}
\caption{\label{STachs} \emph{Number of tachyonic sectors for 54860 type 0 $S$-models, where $k=1,2$ and $i=1,2,3$.}}
\end{table}

As in the case of the $\tilde{S}$-models, we observe a degeneracy in the space of type 0 models and from the sample of 54590 type 0 $S$-models, we found just 89 independent partition functions and the same convergence pattern as shown in Figure \ref{Deg}, therefore we can confidently claim that the lack of tachyon-free type 0 models is a generic result in this class of vacua. It is however worth remembering that our models are defined at a free fermionic point in moduli space and so it may be that such tachyonic instabilities may disappear when a model is translated into a bosonic language and defined away from the fermionic point. The process for doing this in the same basis as we are employing here for the $S$-models is detailed in ref. \cite{FR}.

\section{Misaligned Supersymmetry in Type 0 Models}\label{MSSec}

The partition function of string models encapsulates all the 
information one knows about its structure, symmetries and spectrum.
Thus to fully understand these type 0 models it is essential to get a handle on their partition function. The analysis of the partition function is particularly instrumental in non-supersymmetric constructions, since it gives a complementary tool to count the total number of massless states, and its integration over the fundamental domain corresponds to the cosmological constant. 

The partition function for free fermionic theories is given by the integral
\begin{equation}
    Z = \int_\mathcal{F}\frac{d^2\tau}{\tau_2^2}\, Z_B  \sum_{Sp.Str.} \CC{\alpha}{\beta} \prod_{f} Z \sqbinom{\alpha(f)}{\beta(f)},
    \label{ZInt}
\end{equation}
where $d^2\tau/\tau_2^2$ is the modular invariant measure and $Z_B$ is the bosonic contribution arising from the worldsheet bosons.  The sum over spin structures represents the contributions from the worldsheet fermions. The integral is over the fundamental domain of the modular group 
$$ \mathcal{F} = \{\tau\in\mathbb{C}\,|\,|\tau|^2>1 \;\land\;|\tau_1|<1/2\}. $$
This ensures that only physically inequivalent geometries are counted. The above expression (\ref{ZInt}) specifically represents the one-loop vacuum energy of our theory and so we may refer to it as the cosmological constant.

The practical way to perform this integral is as presented in \cite{D,so10tclass} using the expansion of the $\eta$ and $\theta$ functions in terms of the modular parameter, or more conveniently in terms of $q\equiv e^{2\pi i \tau}$ and $\bar{q}\equiv e^{-2\pi i \bar{\tau}}$. This leads to a series expansion of the one-loop partition function 
\begin{equation}
    Z = \int_\mathcal{F} \frac{d^2\tau}{\tau_2^3} \sum_{n.m} a_{mn} \, q^m \bar{q}^n .
    \label{QPF}
\end{equation}
Writing the partition function as above ensures that the $a_{mn}$ physically represent the difference between bosonic and fermionic degrees of freedom at each mass level, i.e. $a_{mn} = N_b - N_f$.  Further details on the $q$--expansions and the general calculation of these integrals can be found in Section 6 of \cite{so10tclass}.  As expected, on-shell tachyonic states, {\it i.e.} states with $m=n<0$, have an infinite contribution. On the other hand off--shell tachyonic states may contribute a finite value to the partition function. It is also important to note that modular invariance only allows states with $m-n\in \mathbb{Z}$.

In theories with spacetime supersymmetry, it is ensured that the bosonic and fermionic degrees of freedom are exactly matched at each mass level. That is, we necessarily have that $a_{mn} = 0$ for all $m$ and $n$, which in turn causes the vanishing of the cosmological constant as one expects. For our non-supersymmetric models, this level-by-level cancellation is not ensured and so such theories in general produce a non-zero value for $\Lambda$. It is, however, not obviously clear that they should produce finite partition functions, or what form the state degeneracies should take. It has, however, been shown \cite{MSUSY2,MSUSY, carlo}, that the partition functions of non-supersymmetric closed string theoroies possess a special feature called misaligned supersymmetry. 

As one expects, the degeneracy of states grows rapidly going up the infinite tower of massive states. This growth, in theory, could counteract the suppression received from the decreasing contributions from the integrals of (\ref{QPF}) and cause divergences. The mechanism of misaligned supersymmetry, however, causes the states in the massive tower to oscillate between an excess of bosons and an excess of fermions. This behaviour is referred to as boson-fermion oscillation. Instead of cancelling level-by-level as in the supersymmetric case, the cancellation is misaligned causing the oscillation meaning a large positive contribution is followed by an even larger negative contribution and so on.

In the case of type 0 models presented above, due to the presence of on-shell tachyonic states, the partition function (\ref{QPF}) diverges. However, this divergence is contained purely in the tachyonic mode, i.e. the degeneracy of states $a_{mn}$ for $m,n>0$ still behaves in a similar fashion to any other non-tachyonic heterotic theory. This is also emphasised by the presence of misaligned supersymmetry in the massive spectra of our models. Such misalignment of the physical spectrum in heterotic theories is well documented in the literature \cite{MSUSY2,MSUSY,ADM,so10tclass}. It however remains to see whether this mechanism is replicated for type 0 heterotic theories described in previous sections. We find that indeed both sets of type 0 models generated by (\ref{basis}) and (\ref{basisStTi}) exhibit such misalignment of their on-shell massive tower of states. The misalignment pattern appears to have no correlation to whether a specific model has massless fermions or not. The origins of it are, however, not clear in the above cases. 

It was proved in \cite{MSUSY2} that non-SUSY heterotic strings without physical tachyons should always produce such misalignment of their massive spectra. In our case, due to the presence of on-shell tachyons the emergence of misaligned supersymmetry may seem somewhat unexpected and of unknown origin. The analysis presented in the proof of misaligned supersymmetry in \cite{MSUSY2, carlo} relies on modular invariance together with the stated lack of physical tachyons. Since our theory is of course still modular invariant, we can speculate that the emergence of the misalignment should be a consequence of this, however, rigorous analysis is still lacking under these conditions. For the time being, we only present this as an observations for the theories under consideration in this paper, but further investigation may lead to a deeper understanding of the relationship between on-shell tachyons and misaligned SUSY.

\begin{figure}[t]
\centering
\includegraphics[width=0.85\linewidth]{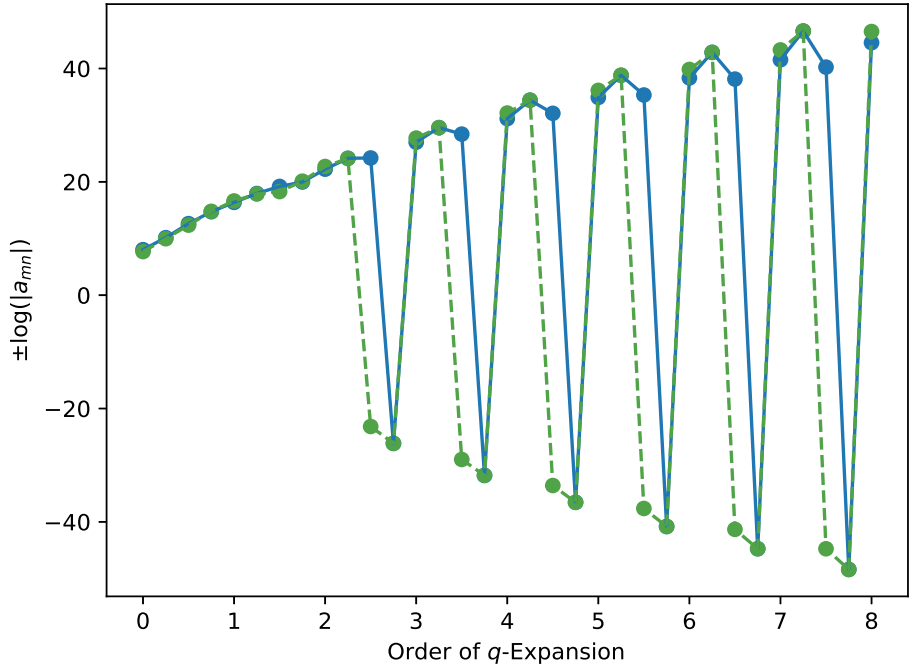}
\caption{\emph{The boson-fermion oscillation of misaligned supersymmetry for the on-shell states of two $\tilde{S}$ models to 8$^{th}$ order in the q-expansion. The overall sign of $\pm\log(\mid a_{mn}\mid)$ is chosen according to the sign of $a_{mn}$.}}
\label{BFOsc}
\end{figure}

As an example, for the $\tilde{S}$-models of Section \ref{StModels} we observe the two general patterns shown in Figure \ref{BFOsc}, or in general a combination of these two.
This is true whether or not the choice of GGSO coefficients project all massless fermions. The only observable difference we find for type 0 models from the partition function point of view is the larger value of  $a_{00}$. This is of course fully expected due to the lack of fermions at the massless level. The misalignment pattern is mostly similar for the $S$--models of Section \ref{SModels}.  It is important to note that, unlike for tachyon--free non--supersymmetric theories, this oscillatory behaviour does not result in a finite value for the cosmological constant due to the presence of the physical tachyons described in Sections \ref{StModels} and \ref{SModels}.

We would like to emphasise that the misaligned supersymmetry found in our
string models is independent of the existence of the tachyon in the spectrum.
In fact, our results demonstrate this fact unequivocally. It is well
known that the absence of a tachyon is sufficient for the existence
of misaligned supersymmetry \cite{MSUSY2,MSUSY, carlo}. However,
our results demonstrate that it is in fact not necessary. It may of course
be the case that the misaligned supersymmetry in vacua that contain tachyons
is only exhibited for special values of the moduli parameters and that moving
away from those points will spoil this property. However, as exhibited
by our results, at least at the free fermionic point, the oscillatory
pattern between fermions and bosons of the massive string spectrum,
persist in vacua that contain tachyons as well. 

\section{Conclusions}\label{conclude}
The $\mathbb{Z}_2\times \mathbb{Z}_2$ heterotic string orbifold produced a rich space of phenomenological
three generation models and serves as a testing ground to explore how the Standard Model
of particle physics may be incorporated in a theory of quantum gravity. The $\mathbb{Z}_2\times \mathbb{Z}_2$ 
orbifold also exhibits a rich symmetry structure which has been of interest from a purely
mathematical point of view. It is of course of much further interest to explore the role
of these mathematical structures in the phenomenological properties of the $\mathbb{Z}_2\times \mathbb{Z}_2$ 
orbifold models. In this paper we explored the existence of type 0 models 
in this class of string
compactifications. Type 0 string vacua are those that do not contain any massless fermionic 
states and have been of interest in other string theory limits, {\it e.g.} the issue of tree
level stability have been studied in the framework of type II orientifolds, whereas 
other authors have advocated that there is holographic duality of the 
type 0B string theory and four dimensional non--supersymmetric Yang--Mills 
theory \cite{glmr}. In this paper we demonstrated the existence of type 0
$\mathbb{Z}_2\times \mathbb{Z}_2$ heterotic string orbifolds. We showed that there exist a large degree 
of redundancy in the space of GGSO projection coefficients when the type 0 restrictions
are implemented. We explored the existence of such configurations in several constructions. 
The one presented in Section \ref{model} correspond to essentially a unique configuration 
out of a priori $2^{21}$ discrete GGSO choices. We demonstrated this uniqueness analytically 
in Section \ref{analyticConds} as well as by the corresponding analysis of the 
partition function in Section \ref{equivalence}.
In Sections \ref{StModels} and \ref{SModels} we performed a wider classification in
$\tilde S$--models and $S$--models, where the first class correspond to compactifications
of a tachyonic ten dimensional heterotic string vacuum, whereas the second correspond to 
compactifications of the ten dimensional non--tachyonic $SO(16)\times SO(16)$. We showed
that the type 0 models in both cases contain physical tachyons at the free fermionic point
in the moduli space. These vacua are therefore necessarily unstable. 
we demonstrated the existence of a misalligned supersymmetry
in the type 0 models that guarantee the finiteness of the one--loop amplitude, aside from 
the divergence due to the tachyonic states. 
Given their rather
restrictive structure, the type 0 models may be instrumental in exploring the string dynamics in
cosmological scenarios in the spirit of the analysis performed in ref. \cite{FR}. Of further
interest will be to explore vacua with a minimal number of bosonic states. Whereas the 
string models always contain massless bosons from the untwisted sector, we may look for
models in which the bosonic states from twisted sectors are projected out. Another
possibility is to look for models in which the fermions from the observable sectors are
balanced by scalars charged under the hidden sector. 

%Note about unstable string vacua like the type 0 ones here perhaps being useful for models of early dynamic/phase transitions in early universe for example
%note about  how tree level stability of type 0 orientifold model achieved in Dan Israel/angelatonj constructions on orientifolds...
% note that the holographic interp of the type 0 theories as large N QCD which may be useful for other systems...little string theory?
%\clearpage

\section*{Acknowledgments}

We would like to that Carlo Angelantonj for fruitful discussions. 
The work of VGM is supported in part by EPSRC grant EP/R513271/1.
The work of BP is supported in part by STFC grant ST/N504130/1.

\bigskip
%\newpage

\bibliographystyle{unsrt}

\end{document}